\documentclass[a4paper]{jpconf}
\usepackage{graphicx}
\usepackage{amsmath,amsthm, amssymb}
\usepackage[english]{babel}
\usepackage{verbatim}
\usepackage{color}
\usepackage{varioref}

\newcommand{\bea}{\begin{eqnarray}}
\newcommand{\eea}{\end{eqnarray}}
\newcommand{\beq}{\begin{equation}}
\newcommand{\eeq}{\end{equation}}

\begin{document}
\title{Analytical approaches to 2D CDT coupled to matter}

\author{Max R Atkin$\,^{a}$ and Stefan Zohren$\,^{b}$$\,^{c}$}

\address{$^a$~Fakult\"{a}t f\"{u}r Physik, Universit\"{a}t Bielefeld, Postfach 100131, D-33501 Bielefeld, Germany}
\address{$^b$ Department of Physics, Pontifica Universidade Cat\'olica do Rio de Janeiro, Rua Marqu\^es de S\~ao Vicente 225, 22451-900 G\'avea, Rio de Janeiro, Brazil}
\address{$^c$ Rudolf Peierls Centre for Theoretical Physics, 1 Keble Road, Oxford OX1 3NP, UK.}

\ead{$^a$ matkin@physik.uni-bielefeld.de, $^b$ zohren@fis.puc-rio.br}

\begin{abstract}
We review some recent results by Ambj{\o}rn et al.\ (1202.4435) and the authors (1202.4322,1203.5034) in which multicritical points of the CDT matrix model were found and in a particular example identified with a hard dimer model. This identification requires solving the combinatorial problem of counting configurations of dimers on CDTs.
\end{abstract}

\section{Introduction}
Casual Dynamical Triangulation (CDT) together with its predecessor Dynamical Triangulation (DT) are approaches to giving meaning to the formal expressions appearing in a path integral quantisation of gravity. The idea is to regulate the path integral by approximating the geometries appearing in the integration by discrete triangulations, thereby replacing the path integral over geometries with a summation over triangulations (s.o.t) each weighted by a discrete form of the original action. Once the s.o.t is computed the regularisation is removed by taking a scaling limit around a critical point described by a continuum theory. 

Much work was done in the early 90s on DT in $D=2,3,4$ dimensions. For $D=2$ the s.o.t could be formulated in terms of a hermitian matrix model \cite{DT}. These models possess critical points whose continuum theory is gravity interacting with matter. For $D>2$ the work was numerical in nature and much less promising; no second order critical point was found at which a continuum theory could be extracted \cite{Bialas:1996wu}. This is due to the preference the theory has for the geometry to undergo spatial topology change.

To remedy this, CDT only sums over triangulations containing a preferred time slicing w.r.t which spatial topology change is forbidden \cite{Ambjorn:1998xu}. For CDT with $D>2$, there are indications of a critical surface on which a continuum theory of extended de-Sitter spacetimes exists (see \cite{review} for a review). There are also interesting results suggesting that in the UV the number of dimensions of the theory reduces to two \cite{Ambjorn:2005db} prompting suggestions that the continuum theory should be identified with a theory known as Horava-Lifshitz gravity \cite{Horava:2009uw}. For $D=2$ the theory without matter has been solved using a variety of methods. In particular, a generalised CDT allowing some topology change has a matrix model formulation \cite{matrix}, opening the possibility of introducing matter and allowing a string worldsheet interpretation; a causal string field theory (CSFT) \cite{Ambjorn:2008ta}.

In principle adding matter to CDT is easy; we simply add extra degrees of freedom to the triangulation. This at least allows numerical investigations to be conducted \cite{Ambjorn:1999gi}. Here however, by ``adding matter'' we mean treating such a matter sector analytically. Of course since no analytical treatments of pure CDT for $D>2$ exists our discussion is confined to two dimensions. There are a number of reasons to be interested in adding matter analytically. Recent numerical work \cite{Ambjorn:2012kd} has indicated that when the central charge of the matter exceeds one, the two-dimensional theory undergoes a transition to a phase with a three-dimensional geometry corresponding to a de-Sitter space similar to that seen in the $D>2$ simulations. Having an analytic understanding of this result would give insight into the transition appearing in the higher dimensional theories \cite{review}. Secondly, due to the lack of matter the CSFT corresponds to strings in zero dimensions. It would be interesting to study CSFT in more complicated backgrounds, hopefully leading to an understanding of how CSFTs differ from usual SFTs \cite{Ishibashi:1993pc}. Indeed there has been recent progress in embedding Horava-Lifshitz gravity into string theory \cite{Griffin:2011xs}. One might suspect that there exists more connections between CDT and string theory and having analytic understanding of matter will be an excellent tool for investigating this.


\section{Multicritical CDT from a matrix model}
The matrix model approach to (C)DT is based on the fact that triangulations are in bijection \cite{DT} with Feynman diagrams of matrix models of the form,
\beq
\label{ZMM}
Z = \int [d\phi] e^{-N\Tr V(\phi)} = \int^\infty_{-\infty} \prod^N_{i=1} d \lambda_i \prod_{j \neq i} (\lambda_j -\lambda_i)^2 e^{-N \sum^N_{i=1} V(\lambda_i)},
\eeq
where $\phi$ is an $N\times N$ hermitian matrix, $[d\phi]$ is the standard measure on such matrices and $V(\phi) = \sum^\infty_{k=0} t_k \phi^k/k$ is known as the potential. We therefore identify $Z$ with the s.o.t with an action determined by the $t_k$. The advantage of the matrix model approach is that well developed techniques for computing $Z$ exist \cite{DT}. Indeed, in the second equality we have written the integral in terms of the eigenvalues $\lambda_i$ of $\phi$. Such an integral can be viewed as a gas of charged particles in a potential and may be evaluated by orthogonal polynomial methods. Once $Z$ is computed, we take a scaling limit about a second order phase transition to recover a continuum theory. Such a transition occurs when the eigenvalues of $\phi$ begin to spill over a confining barrier. By tuning the parameters appearing in $V$ one can obtain critical points describing the continuum limit of DT coupled to $(2,2p+1)$ conformal minimal matter \cite{DT}.

In the last few year it was realised \cite{matrix} that pure CDT may also be obtained as a scaling limit of the matrix model \eqref{ZMM} if one uses the potential $V_\mathrm{CDT}(\phi) = \frac{1}{\beta}\left(-g \phi + \phi^2/2 - g \phi^3/3 \right)$. In \cite{matrix} it is shown that $\beta$ acts as a coupling for spatial topology change. To forbid topology change we require $\beta = 0$, which is clearly singular in the above action, partially explaining why it took so long to find a matrix model for CDT. The insight of \cite{matrix} was to make an scaling ansatz for the variables of, 
\beq
\label{scaleansatz}
g = g^* - a^2 \Lambda \qquad \mathrm{and} \qquad \beta = a^3 g_s,
\eeq
where $a$ is lattice length which goes to zero in the scaling limit and hence $\beta \rightarrow 0$. This leads to a {\emph{continuum theory}} in which there is a coupling for spatial topology change. By moving the term $(\lambda_j -\lambda_i)^2$ in \eqref{ZMM} in to the potential we see that $\beta$ controls the strength of repulsion between eigenvalues. By scaling $\beta$ to zero we allow the eigenvalues to collect at the bottom of the potential. A similar transition, in which eigenvalues {\emph{appear}} in a new minima of the potential, was studied in \cite{cutbirth} in which the scaling limit about the point where the eigenvalues first appear is again described by a matrix model. Here, we find a similar phenomenon if we scale $\phi$ in \eqref{ZMM} as $\phi  = \phi^* + a \Phi$. This results in the scaled partition function \cite{matrix},
\beq
\label{scaledZMM}
\tilde{Z}_\mathrm{CDT} = \int [d\Phi] e^{-\frac{N}{g_s} \left(2 \Lambda \Phi - \frac{1}{6} \Phi^3 \right)}.
\eeq
We stress here that one can compute continuum CDT amplitudes from \eqref{scaledZMM} or by first computing the discrete amplitude using \eqref{ZMM} and then taking the limit with \eqref{scaleansatz}. To consider matter one follows DT and looks for a non-trivial scaling limit for higher order $V$ in which $\beta \rightarrow 0$. Following \cite{matrix}, in \cite{multi} and later \cite{higher} it was realised that by choosing the potential and scaling ansatz to be of the form,
\bea
&&V_\mathrm{mCDT}(\phi) = -g \phi + \frac{1}{2} \phi^2 - g \sum^{m+1}_{k=1} \frac{t_k}{k} \phi^k \\
\label{kscaleansatz}
&&\phi = \phi^* + a \Phi, \qquad g = g^* - a^m \Lambda \qquad \mathrm{and} \qquad \beta = a^{m+1} g_s,
\eea
we obtain the scaled partition function,
\beq
\label{contZm}
\tilde{Z}_\mathrm{mCDT} = \int [d\Phi] e^{-\frac{N}{g_s} \mathrm{Tr} \left[(-\partial_g \partial_\phi V^*) \Lambda \Phi + \frac{1}{(m+1)!}(\partial^{m+1}_\phi V^*) \Phi^{m+1} \right] }.
\eeq
The critical point $(\{t^*_k\}, g^*, \phi^*)$ is defined such that $V^{(n)}(\{t^*_k\}, g^*, \phi^*) = 0$ for $n \leq m$. There are a number of interesting observations about \eqref{contZm}. Firstly, the continuum theory is again a matrix model; this should be compared with \cite{cutbirth} and it would be interesting to understand the relation between these two results. Note also that the scaling limit does not change $N$ and hence $N$ is a free parameter. It was argued for the pure case $m=2$ \cite{matrix} and for $m>2$ \cite{higher}, that $N$ can be identified with a coupling that distinguishes the splitting and joining of universes {\emph{in the continuum theory}}. If we do not wish to distinguish these processes we can set $N = 1$ thereby giving a simple expression for the entire summation of the topological expansion. Finally, we identify \eqref{contZm} as a degenerate form of the generalised Kontsevich model. Such a model has appeared in non-critical string theory before as the partition function of the continuum theory of a stack of FZZT branes \cite{Hashimoto:2005bf}. The continuum limit of $m$th order multi-critical CDT is therefore equivalent, at least mathematically, to this. Such an observation was also made in \cite{Ambjorn:2009rv} for the case of pure CDT. It would of course be highly interesting to understand if there exists a physical explanation for this coincidence. 

\section{Identification of $m=3$ with hard dimers}
\begin{figure}[t]
\centering 
\includegraphics[scale=0.33]{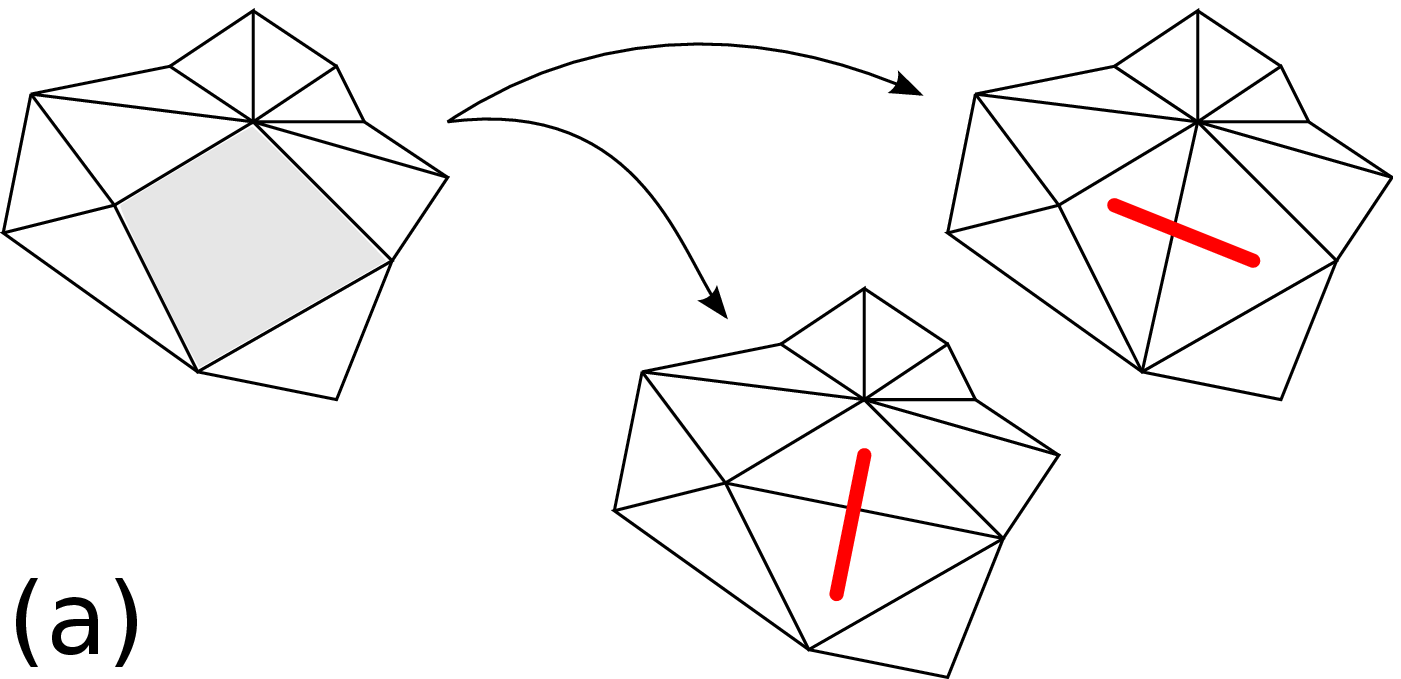}
\quad\quad \quad\quad 
\includegraphics[scale=0.32]{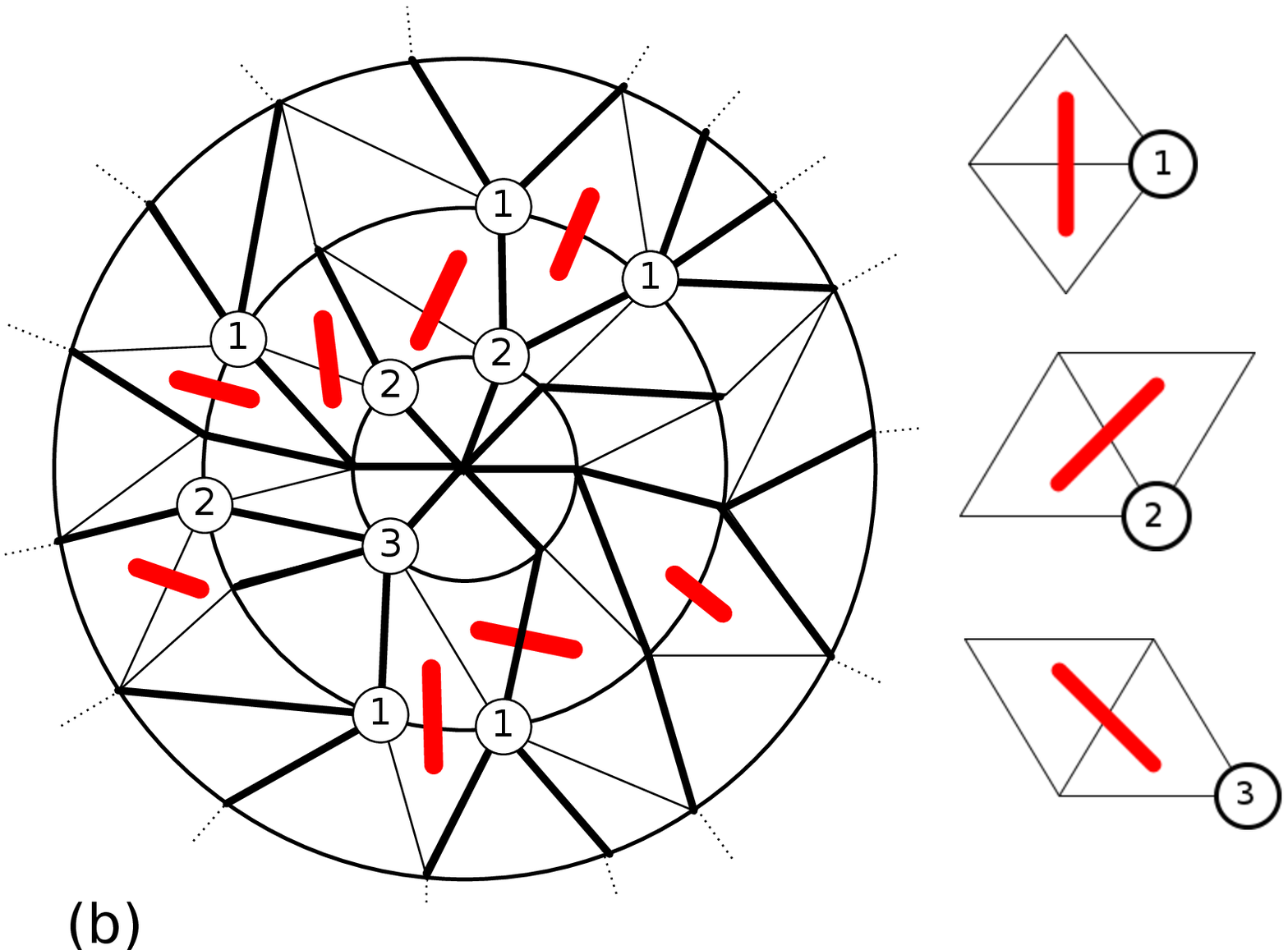}
\caption{{\bf (a)} The mapping between squares and hard dimers. This mapping does not respect the causal structure of CDT. {\bf (b)} A bijection between hard dimers in CDT and labelled trees. The tree is shown in thickened lines and the labels are added using the rules on the right.}
\label{CDTbijection} 
\end{figure}

Identifying the continuum theory for $m>2$ is difficult as we have no continuum formulation of the gravity sector. This is in contrast with DT in which the continuum gravity sector is Liouville theory and for multicritical points it is known the continuum theory is a tensor product of Liouville theory with a minimal model. Progress on this problem was made in \cite{multi,dimer} where $m=3$ was identified with a hard dimer model as is also the case for DT with a quartic interaction. However in DT it is clear how hard dimers, which are objects occupying two adjacent triangles such that no two dimers occupy the same triangle, can be mapped to squares; see Figure \ref{CDTbijection}(a). For CDT such a mapping is no longer valid.

For $D=2$ the partition function is the generating function for CDTs of a given size containing a given number of dimers. The idea of \cite{dimer} was to explicitly solve a hard dimer model on CDT by use of a bijection between CDT and labelled tree graphs; this is shown in Figure \ref{CDTbijection}(b). Unlabelled trees are easy to count since any sub-graph is also a tree, allowing a recursive equation for the generating function of the number of trees of a given size to be found and solved. Counting the labelled trees in correspondence with hard dimers is a difficult task; the bijection removes links and hence local interactions in the CDT can become non-local in the tree, thereby preventing the trees from being defined recursively.

In \cite{dimer} enough non-local interactions were removed to allow an analytical treatment but not so many as to prevent new critical behaviour. This can be seen in Figure \ref{CDTbijection} (a); we only allow a restricted class of dimers to appear. Note that the type-3 dimers are non-local, however they can be accounted for by noting there exists a bijection between trees containing such dimers and distinct trees in which they are replace by type-2 dimers (see \cite{dimer} for details). Dimers of type one and two only affect dimers which are adjacent to them in the tree. Hence trees containing only these can be counted recursively. The critical exponents of the model were computed in \cite{dimer} and found to agree with the values predicted from the matrix model.

\section{Summary}
We have given an overview of the recent progress in analytically analysing matter coupled to CDT together with a number of examples of why such work is important. We have seen that it is a useful tool for connecting CDT with string theory and will hopefully help in clarifying the manner in which the de-Sitter phase of higher dimensional CDT arises. A natural continuation of this work is to generalise the results here to any $(p,q)$ minimal model with the first target being the Ising model. The unitary models are particularly interesting as one may compare to the numerical work of \cite{Ambjorn:1999gi}.

\end{document}